\documentclass[conference,twocolumn,10pt,twoside]{IEEEtran}

\bstctlcite{IEEEexample:BSTcontrol}

\normalsize
\usepackage{amssymb,amsthm}
\usepackage{cite}
\usepackage{tabularx}

\newtheorem{rem}{Linearization Technique}

\ifCLASSINFOpdf
   \usepackage[pdftex]{graphicx}
\else
   \usepackage[dvips]{graphicx}
\fi

\usepackage[cmex10]{amsmath}

\usepackage{algorithm}
\usepackage{algpseudocode}

\usepackage{stfloats}
\usepackage[caption=false,font=footnotesize]{subfig}
\begin{document}
\title{Optimal User Association for Massive MIMO Empowered Ultra-Dense Wireless Networks}

\author{\IEEEauthorblockN{Antonis G. Gotsis, Stelios Stefanatos, and Angeliki Alexiou}\\
\IEEEauthorblockA{Department of Digital Systems, University of Piraeus, Greece\\
%150 Androutsou Street, Room 303, GR18532, Piraeus, Greece\\
Email: \{agotsis, sstefanatos, alexiou\}@unipi.gr}}

\markboth{}%
{}



\maketitle

\begin{abstract}
\boldmath
Ultra network densification and Massive MIMO are considered major 5G enablers since they promise huge capacity gains by exploiting proximity, spectral and spatial reuse benefits. Both approaches rely on increasing the number of access elements per user, either through deploying more access nodes over an area or increasing the number of antenna elements per access node. At the network-level, optimal user-association for a densely and randomly deployed network of Massive MIMO empowered access nodes must account for both channel and load conditions. In this paper we formulate this complex problem, report its computationally intractability and reformulate it to a plausible form, amenable to acquire a global optimal solution with reasonable complexity. We apply the proposed optimization model to typical ultra-dense outdoor small-cell setups and demonstrate: (i) the significant impact of optimal user-association to the achieved rate levels compared to a baseline strategy, and (ii) the optimality of alternative network access element deployment strategies. 
\end{abstract}


%\newpage

% \fi
\IEEEpeerreviewmaketitle

\section{Introduction}\label{sec:Intro}
5G wireless networks target the provision of a sustainable solution to the radio capacity crunch predicted over the next decade~\cite{NGMN5GWPExec14}. Recent studies attempting to specify future network requirements state that 5G networks should support multiple orders of magnitude increased network and per user capacity levels, compared to today~\cite{METISD11}. Although we are at the very beginning of defining a concrete 5G system concept, there are already strong indications for how the 5G enabling technologies portfolio will be formed~\cite{NGMN5GWPExec14,METISD62,AnBu14,LiNi14,EXALTEDD24}. Following recent progress, we identify and focus on two key 5G evolution solutions:
\begin{itemize}[leftmargin=10pt]
\item Ultra Network Densification, namely multiple orders of magnitude higher infrastructure density levels compared to today's macro-cell networks \cite{HwSo13}. Ultra-Dense Networks (UDNs) will be characterized by comparable serving access nodes (small-cells) and served nodes (user devices) densities, up to the extreme point that each user is served by its own access node, exploiting the associated proximity and spectral reuse benefits.
\item Massive MIMO, namely empowering today's base stations with multiple orders of magnitude greater number (tens or even hundreds) of antenna elements compared to conventional MIMO sizes (4 or 8 antennas per base station which is the case for LTE/LTE-A), surpassing the number of active users~\cite{LaEd14}. Massive MIMO technology offers both spectral efficiency and resources reuse benefits due to the exploitation of massively available spatial degrees of freedom.
\end{itemize} 

Interestingly both technologies are built on the same design principle, that is the over-provisioning of infrastructure access elements per user. However, their focus is different, since UDNs tend to create highly localized access areas with improved channel conditions due to link distance reduction, whereas Massive MIMO leverages powerful spatial processing to create multiple interference-free user-access node links. 

From a network-level perspective, the problem of selecting the best serving access node for each user, usually referred as user-association, is highly challenging in UDNs  due to random topology deployment and significant load variations observed in different access nodes~\cite{HoRa14}. The problem becomes further exacerbated when Massive MIMO comes into play, since the presence of excess antenna elements per access node calls for an optimal exploitation of the available spatial degrees of freedom~\cite{BeBu14}. In this paper we formulate and solve, for the first time to the best of our knowledge, the optimal user association problem for Massive MIMO empowered homogeneous ultra-dense networks. We leverage the optimal solution to comment on the interplay of Massive MIMO and UDN technologies, and explore the performance trends as a function of access node and antenna element densities in future network setups.

\subsection{Related Work}
The potential of infrastructure densification for achieving 5G capacity targets was recently demonstrated in~\cite{METISD63}. Results based on simulation campaigns for indoor environments, revealed that UDN deployments corresponding to a single access node located at every room and serving at most three user devices, are able to support multiple orders of magnitude higher user rates, when combined with a moderate spectrum increase. The importance of optimizing user-association in UDNs has been stressed in the context of 3GPP small-cell enhancements study item~\cite{3GPP_36872,3GPP_R1133246}. It has been demonstrated through system-level simulations that significant performance improvements could be achieved when considering both channel and load conditions during user-association decision. Nevertheless, only SISO network setups were considered and the applied user-association criteria were based on heuristic rules, rendering their optimality questionable. Interesting initial results for optimizing user-association in a limited 2x2 (2 access nodes and 2 users) SISO network setup have been recently reported in \cite{WiOs15}; however it is not clear how these findings could be generalized in scenarios involving more nodes and when the access nodes operate in the Massive MIMO regime.

A second line of works has recently emerged, striving to answer the ``antenna element deployment dilemma". The authors in~\cite{DiAm14,HoYu14} explored which deployment strategy is optimal given an excess number of antenna elements over a serving area: to concentrate all elements on a single base station or to spread multiple single-antenna access nodes over the area. No single winner was identified, since conflicting performance trends were reported, depending on the considered environment type (indoor or outdoor). Differently from our work, these studies did not consider the potential of joint channel- and load-based user-association, since each user was assumed to connect to the closest access node. The most relevant work to ours is \cite{BeBu14}. The authors first provide an approximate, yet fairly accurate expression, abstracting Massive MIMO impact on the achieved rate performance, in multi-cell setups. Then, based on this expression, they consider optimal load balancing strategies realized over a finite time horizon, for Massive MIMO HetNets.

\subsection{Contributions}
In this paper, building upon the Massive MIMO abstraction model of \cite{BeBu14} we:
\begin{itemize}[leftmargin=10pt]
\item Formulate the problem of user-association for maximizing the guaranteed user rate across a Massive MIMO empowered ultra-dense network; 
\item Show that problem is computationally intractable for practical network cluster sizes and demonstrate that it admits an exact reformulation to a plausible form, which allow us to acquire the global optimal solution with reasonable complexity;
\item Apply the transformed optimization model to acquire the optimal associations and achieved performance levels for a representative future network setup, considered by 3GPP;
\item Leverage the acquired system-level results to quantify the impact of densification and Massive MIMO size enhancements to future networks, as well as explore the optimal strategy for deploying a given amount of antenna elements across the network, namely \emph{concentrate} elements to fewer access nodes or \emph{distribute} elements to a larger number of nodes;
%\item Conclude that: (1) optimal user-association significantly enhances the achieved rate performance levels with reduced number of access nodes active, compared to a baseline channel-based only strategy; (2) for a given amount of antenna elements deployed across a network, the optimal distribution strategy depends on the overall transmission power budget.
\end{itemize} 

The rest of this paper is organized as follows. Sec.~\ref{sec:SystemModel} describes the system model, while Sec.~\ref{sec:OPT} introduces the user-association problem, provides its mathematical formulation, and the steps for transforming it to a more plausible form. Sec.~\ref{sec:Results} reports on numerical results and discuss key findings, whereas Sec.~\ref{sec:Concl} concludes the work.

%\cite{AnBu14,LiNi14,NGMN5GWPExec14,SKTWP14, HwSo13, 4GAmer5GSummaryJun14, 4GAmer5GReqAndSolOct14, PoMo14, HoRa14,METISD63,,BeBu14,DiAm14,HoYu14,BjLa14}
%
%\cite{CVX}
%\cite{GUROBI}
%
%\cite{3GPP_36872}

%\begin{figure}
%\centering
%\resizebox{13cm}{!}{\includegraphics{UDN_mag_Fig_1}}
%\caption{An ultra-dense network infrastructure composed by operator- and user-deployed heterogeneous serving access nodes, multiple types of user and machine served nodes, and disruptive devices acting as prosumers.}
%\label{UDN_mag_Fig_1}
%\end{figure}
%
%\begin{table}
%\renewcommand{\arraystretch}{1.3}
%\caption{Densification Savings Provided by Coordinated Policies}
%\label{table_example}
%\centering
%\begin{tabular}{ |p{8em}||p{5em}|p{5em}| }
%\hline
%\multirow{2}{10em}{guaranteed UE rate (bps/Hz)}  &\multicolumn{2}{|c|}{\% savings in densification ratio} \\ \cline{2-3}
%& Policy I & Policy II \\ \hline \hline
%0.1 & 60 \% & 86 \%  \\
%0.5 & 45 \%  & 82 \%  \\
%1 & 41 \%  & 81 \%  \\
%\hline
%\end{tabular}
%\end{table}

\section{System Model}\label{sec:SystemModel}
We consider a set of $\left| \mathcal{M} \right| = M$ densely and randomly deployed multi-antenna access nodes (ANs) serving $\left| \mathcal{K} \right| = K$ single-antenna users (UEs), similar to the overlay outdoor deployment scenario assumed in the context of 3GPP Release 12 small-cell densification theme~\cite{3GPP_36872}. We assume a single cluster of ANs for simplicity, and an over-provisioning of access elements per UE. In particular: i) we focus on the \textit{ultra-dense network} (UDN) region, meaning that the number of ANs and UEs is comparable ($K \approx M$)~\cite{TuLi14}; ii) consider operation in the \textit{Massive MIMO} (MM) regime, meaning that each AN $m$  is equipped with a very large number of antenna elements, let $L$, at least an order of magnitude greater than the number of UEs it accommodates ($L \gg {S_m}$, where $S_m$ is the number of UEs served by AN $m$), such that all UEs are served simultaneously over the same time-frequency resource unit~\cite{LaEd14}. It is also assumed that each UE is served by a single AN and no inter-AN coordination in the data plane is supported.

For a homogeneous network deployment scenario, it is reasonable to assume a common number of antenna elements per AN, $L$, and a common transmission power level per AN, $p$. We define the channel state information (CSI) factors  for all possible UE-AN combinations, including the large-scale distance-based path-gain normalized to a given noise density level ($g_{km}$) and small-scale rayleigh fading. Assuming user $k^*$ served by AN $m^*$, then according to the recent work of \cite{BeBu14}, the achieved spectral efficiency (or simply rate) is  given by $r = {\log _2}\left( {1 + {\gamma _{{k^*}{m^*}}}} \right)$, where ${{\gamma _{{k^*}{m^*}}}}$ is the \emph{effective} SINR when the system is operating in the Massive MIMO region. The latter is well approximated by \cite{BeBu14}:
\begin{equation}\label{eq:SINR-MM}
{\gamma _{{k^*}{m^*}}} = \left( {\dfrac{{L - {S_{{m^*}}} + 1}}{{{S_{{m^*}}}}}} \right) \cdot \dfrac{{p \cdot {g_{{k^*}{m^*}}}}}{{1 + \sum\limits_{m \ne {m^*}} {p \cdot {g_{{k^*}m}}} }}. 
\end{equation}
We report the major relevant observations of \cite{BeBu14} for convenience: first the achieved SINR and equivalently the spectral efficiency or simply rate, does not depend on the small-scale channel fading (notice that in \eqref{eq:SINR-MM} only the large-scale factors are present), and second, each UE's rate depends on the number of UEs accommodated by its serving AN. Different from \cite{BeBu14}, we assume a constant activity factor for each UE, namely all UEs inside each cluster are served simultaneously. Note that \eqref{eq:SINR-MM} is composed of:
\begin{itemize}[leftmargin=10pt]
\item the factor ${{p \cdot {g_{{k^*}{m^*}}}} \mathord{\left/
 {\vphantom {{p \cdot {g_{{k^*}{m^*}}}} {\left( {1 + \sum\limits_{m \ne {m^*}} {p \cdot {g_{{k^*}m}}} } \right)}}} \right.
 \kern-\nulldelimiterspace} {\left( {1 + \sum\limits_{m \ne {m^*}} {p \cdot {g_{{k^*}m}}} } \right)}}$, which is equivalent to the standard SINR formula for single-antenna networks, when accounting only for large-scale channel effects;
\item the factor ${{\left( {L - {S_{{m^*}}} + 1} \right)} \mathord{\left/
 {\vphantom {{\left( {L - {S_{{m^*}}} + 1} \right)} {{S_{{m^*}}}}}} \right.
 \kern-\nulldelimiterspace} {{S_{{m^*}}}}}$, which abstracts the Massive MIMO gain, compared to a reference conventional single-antenna setup.
\end{itemize}
In this paper, we search for identifying the optimal user-to-AN association decision which maximizes the worse rate level across all UEs, assuming that the network CSI information is perfectly known. The selected optimization criterion provides in essence the minimum guaranteed performance level, ensuring a fair handling of the various UEs, contrary to a sum-rate maximization criterion, which could resort to highly unbalanced achieved rates on a UE basis\footnote{Note that one of the key requirement for 5G is the provision of uniform quality-of-service levels (see for example \cite{METISD11}).}.

\begin{figure*}[!t]
\centerline{
\subfloat[{Baseline Scheme}]
{\includegraphics[scale=0.45]{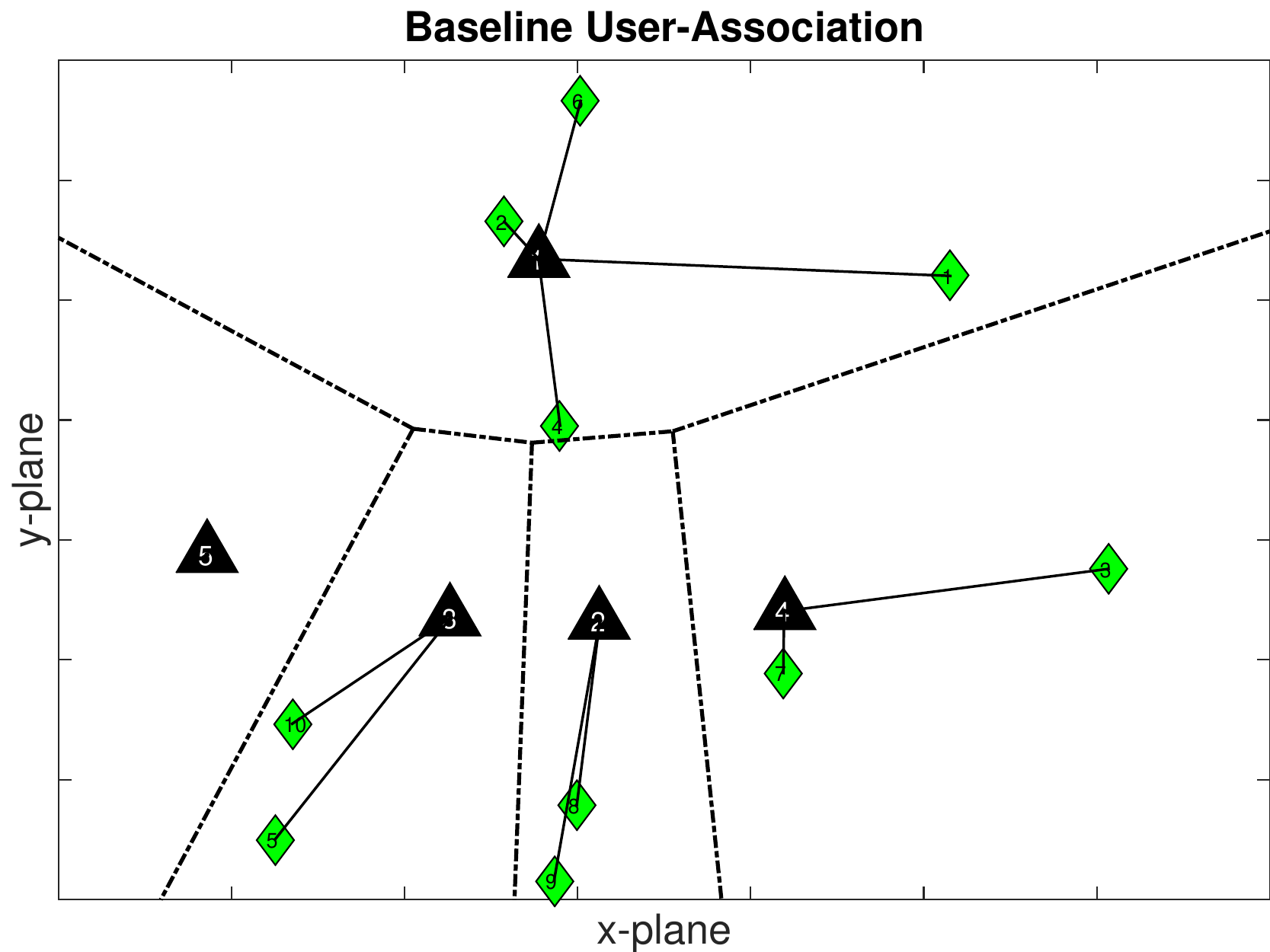}
\label{fig:SystemSnapshot_VOR}}
\hfill
\subfloat[{Optimal Scheme}]
{\includegraphics[scale=0.45]{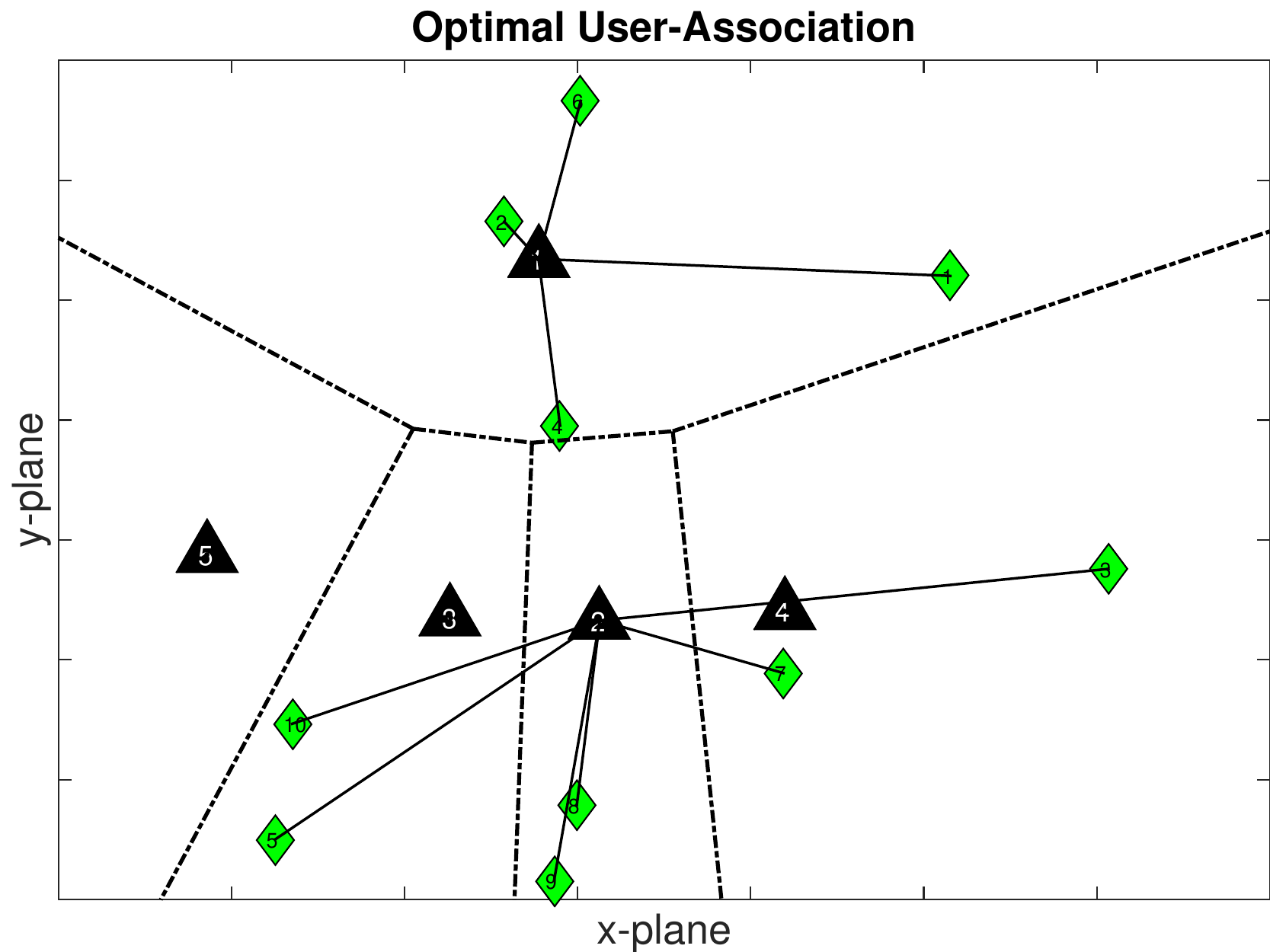}
\label{fig:SystemSnapshot_OPT}}
}
\caption{A user association example for a random topology system snapshot involving a single cluster of 5 access nodes (each AN represented by a $\bigtriangleup$) and 10 users (each UE represented by a $\Diamond$). Dashed lines depict the coverage limits for each AN considering only geographical information. Solid lines represent the user association decisions.  Fig.~\ref{fig:SystemSnapshot_VOR} corresponds to the baseline user-association scheme, where each UE is served by its closest AN, providing the strongest received useful power. Fig.~\ref{fig:SystemSnapshot_OPT} corresponds to the optimal user-association decision. Notice that compared to the baseline scheme, the optimal scheme concentrates users served by ANs 2,3 and 4 to a single AN (2), since deactivating ANs 3 and 4 reduces network interference. For this example, optimal user-association enhances rate performance at a lower access node utilization level.}
\label{fig:SystemSnapshot}
\end{figure*}

\section{User Association In Massive MIMO UDNs: Problem Formulation and Optimal Solution}\label{sec:OPT}

\subsection{Motivation - Problem Statement}\label{sec:OPT_a}
In traditional cellular networks the density of UEs is usually multiple orders of magnitude greater than that of ANs, and each UE is typically associated with the AN providing the best channel conditions (in terms of received useful power). This results in having non-idle ANs, balanced traffic loads offered per AN, and a uniform interference landscape. In Massive MIMO empowered UDNs (MM-UDNs) the following differentiating factors are noted:
\begin{itemize}[leftmargin=10pt]
\item The densities of ANs and UEs are comparable, hence multiple potentially serving ANs lie at the vicinity of each UE, with good channel conditions;
\item The traffic load distribution among different ANs is highly unbalanced, opening up opportunities for dynamic load coordination among neighboring ANs;
\item The interference landscape depends heavily on user association decisions; for example one could de-activate an AN serving a limited number of UEs (see for example in \cite{3GPP_R1133246}), shift these UEs to a neighbor already active AN, thus reducing network interference. 
\item The excess number of available (spatial) degrees of freedom which could be utilized for enhancing useful received power and/or multiplexing more users on a per-AN basis, further perplex the user-association optimization decision making.
\end{itemize}

Based on the above observations we argue that in future networks, user association should be intelligently configured on a network basis, striking a balance between the following trade-offs:
\begin{itemize}[leftmargin=10pt]
\item \textit{Maximizing useful received signal power} (by selecting the most proximal AN) vs \textit{Maximizing received SINR} taking into account the capability to also control network interference. An example is shown in Fig.~\ref{fig:SystemSnapshot}, where UEs served by ANs 3 and 4 according to the baseline association scheme (Fig.\ref{fig:SystemSnapshot_VOR}), should be shifted to the farther AN-2  (allowing ANs 3 and 4 to be de-activated), according to the optimal association scheme (Fig.\ref{fig:SystemSnapshot_OPT}).
\item \textit{Load balancing} (by distributing the UE population to multiple ANs) vs \textit{Load concentration} (by shifting UEs to a subset of (or even a single) Active AN), towards exploiting the spatial and spectral degrees of freedom, and in essence optimally tuning the Massive MIMO gain factor given in the effective SINR formula \eqref{eq:SINR-MM}. For example in the configuration of Fig.\ref{fig:SystemSnapshot_OPT} it is not optimal to shift all UEs to a single-AN, since the Massive MIMO gain will become too limited.
\end{itemize}

\subsection{Problem Formulation}
We introduce:
\begin{itemize}[leftmargin=10pt]
\item a set of binary variables ${\left\{ {{\alpha _{km}}} \right\}_{k \in \mathcal{K},m \in \mathcal{M}}} \in \left\{ {0,1} \right\}$, reflecting the potential UE-to-AN associations, where $\alpha_{km} = 1$ if UE $k$ is associated with AN $m$; 
\item a set of binary variables ${\left\{ {{\rho _m}} \right\}_{m \in \mathcal{M}}} \in \left\{ {0,1} \right\}$, indicating the activity of each AN, where $\rho_{m} = 1$ if AN $m$ is active.
\end{itemize}
An AN is considered active if at least one UE is associated with it, else is turned off to avoid creating unnecessary interference. The following constraint expressions clearly hold:
\begin{subequations}
\noindent\begin{tabular}{@{}*{2}{m{0.5\linewidth}@{}}}
\begin{equation}
\sum\limits_m {{\alpha _{km}}}  = 1,\forall k, \label{eq:OPTa}
\end{equation} &
\begin{equation}
   {\alpha _{km}} \leqslant {\rho _m},\forall k, \label{eq:OPTb}
\end{equation}
\end{tabular}
\end{subequations}
where \eqref{eq:OPTa} guarantees that each UE should be associated with a single AN, and \eqref{eq:OPTb} that for an inactive AN, no UE could be associated with it. Utilizing the user-association variables, the number of UEs served by an AN $m$ is simply given by ${S_m} = \sum\limits_{i \in \mathcal{K}} {{\alpha _{im}}}$. Introducing the decision variables to \eqref{eq:SINR-MM}, after a simple manipulation we may express the achieved effective SINR for an arbitrary UE-AN combination as:
\begin{equation}\label{eq:SINR-MM-OPT}
{\gamma _{km}} = \dfrac{{{g_{km}} \cdot \left( {L + 1} \right) - {g_{km}} \cdot \sum\limits_i {{\alpha _{im}}} }}{{\left( {{1 \mathord{\left/
 {\vphantom {1 p}} \right.
 \kern-\nulldelimiterspace} p}} \right) \cdot \sum\limits_i {{\alpha _{im}}}  + \sum\limits_{i} {\sum\limits_{m' \ne m} {{g_{km'}} \cdot {\alpha _{im \cdot }} \cdot } {\rho _{m'}}} }}.
\end{equation}
The above fractional expression reveals the impact of different parameters on the achievable per-UE performance:
\begin{itemize}[leftmargin=10pt]
 \item The numerator is optimized by assigning each UE to its most proximal AN, so as to maximize the experienced path-gain $g_{km}$, given that $L \gg \sum\limits_i {{\alpha _{im}}}$ holds in the Massive MIMO regime;
 \item The denominator is optimized by minimizing the number of co-served UEs (expressed through $\sum\limits_i {{\alpha _{im}}}$) and the number of interfering ANs (indicated by the $\rho _{m'}$ variables that equal unity).
 \end{itemize} 
We aim at finding the user-association decision which corresponds to the maximum effective SINR guaranteed for all UEs (equivalently the max-min SINR). This corresponds to solving the following optimization problem:
\begin{equation}\label{eq:OPT-PF}
\left\{
 \begin{aligned}
\underset{\theta,\{\alpha _{km}\},\{\rho _m\}}{\text{maximize}}& \theta \\
\text{subject to } &	{\gamma _{km}} \geqslant {\alpha _{km}} \cdot \theta ,\forall k,\forall m ,\eqref{eq:OPTa}, \eqref{eq:OPTb} , \eqref{eq:SINR-MM-OPT}
 \end{aligned}
\right\}
\end{equation}
where $\theta$ is the maximum guaranteed effective SINR. The constraint expression ${\gamma _{km}} \geqslant {\alpha _{km}} \cdot \theta$ is interpreted as follows: if the UE $k$ is served by the AN $m$, namely $\alpha_{km}=1$, then the achieved SINR should be at least equal to the guaranteed common level $\theta$, namely ${\gamma _{km}} \geqslant \theta$, whereas if not, namely $\alpha_{km}=0$, it is irrelevant, namely ${\gamma _{km}} \geqslant 0$.

With regard to the characterization of the optimization problem stated in \eqref{eq:OPT-PF} we observe that:
\begin{itemize}[leftmargin=10pt]
\item Both discrete (binary) and continuous variables are involved;
\item The achieved SINR given in \eqref{eq:SINR-MM-OPT} corresponds to a (non-linear) fractional expression of binary (numerator) and products of binary (denominator) variables;
\item The minimum SINR constraint expression involves the (non-linear) product of a binary and a continuous variable.
\end{itemize}
Therefore the problem in \eqref{eq:OPT-PF} is a mixed integer non-linear program (MINLP), which even for small dimensions, is extremely difficult to solve while providing global optimality guarantees~\cite{LeLe12}. 

\subsection{Reformulation towards acquiring the optimal solution}
To cope with the intractability of formulation \eqref{eq:OPT-PF}, we will demonstrate that the problem admits an exact linearization-reformulation, allowing us to transform it to a mixed integer linear program (MILP). The advantage of MILPs is that there exist powerful methods and tools for acquiring their global optimal solutions, at least for practical cluster sizes involving one to two tens of ANs and UEs.

Notice that the constraints in \eqref{eq:OPTa} and \eqref{eq:OPTb} are already linear with respect to $\left\{ {{\alpha _{km}}} \right\}$ and $\left\{ {{\rho _m}} \right\}$ variables. Next, by incorporating \eqref{eq:SINR-MM-OPT} into \eqref{eq:OPT-PF} we get:
\begin{equation}\label{eq:OPT-PF-MinSINRconFull}
\dfrac{{{g_{km}} \cdot \left( {L + 1} \right) - {g_{km}} \cdot \sum\limits_i {{\alpha _{im}}} }}{{\left( {{1 \mathord{\left/
 {\vphantom {1 p}} \right.
 \kern-\nulldelimiterspace} p}} \right) \cdot \sum\limits_i {{\alpha _{im}}}  + \sum\limits_{i} {\sum\limits_{m' \ne m} {{g_{km'}} \cdot {\alpha _{im \cdot }} \cdot } {\rho _{m'}}} }} \geqslant {\alpha _{km}} \cdot \theta ,\forall k,\forall m.
\end{equation}
Although \eqref{eq:OPT-PF-MinSINRconFull} involves multiple non-linear terms, we will show how this expression could be exactly linearized.
Before introducing the exact reformulation procedure, we present two common reformulation techniques, answered in related mathematical programming literature, which deal with products involving two binary variables or a binary and a continuous variable respectively~\cite{ChBa10,Wu97}. The validity of these techniques is proven by simple inspection.
\begin{rem}\label{rem:BBP}
A product of two binary variables $x$ and $y$ can be replaced by a new auxiliary binary variable $z = xy$, along with a set of three linear constraint expressions: $z \leqslant x$, $z \leqslant y$, and $z \geqslant x + y - 1$.
\end{rem}
\begin{rem}\label{rem:BCP}
A product of a binary variable $x$ and a continuous positive variable $y$ can be replaced by a new continuous auxiliary variable $z = xy$, along with a set of four linear constraint expressions: $y-z \leqslant K_y(1-x)$, $z \leqslant y$, $z \leqslant K_yx$, and $z \geqslant 0$, where $K_y$ is large number guaranteed to be greater than the maximum value that $y$ could take.
\end{rem}
In what follows we will show how each product term appearing in \eqref{eq:OPT-PF-MinSINRconFull} can be linearized, using the above techniques. 
\paragraph{Step 1} With respect to the left-hand side of \eqref{eq:OPT-PF-MinSINRconFull}, the only non-linear term is the double-summation expression of the denominator, which involves products of two binary variables. Following Technique~\ref{rem:BBP}, we introduce the set of auxiliary binary variables ${\left\{ {{z_{imj}}} \right\}_{i \in K,m \in M,j \in M\backslash \left\{ m \right\}}}$, through which we can replace the non-linear terms ${\sum\limits_i {\sum\limits_{m' \ne m} {{g_{km'}} \cdot {\alpha _{im \cdot }} \cdot } {\rho _{m'}}} }$ with the linear terms $\sum\limits_i {\sum\limits_{j \ne m} {{g_{kj}} \cdot {z_{imj}}} } $ along with the following set of linear constraint expressions:
\begin{equation}\label{eq:AUX1}
\begin{gathered}
  {z_{imj}} \leqslant {\alpha _{im}},{z_{imj}} \leqslant {\rho _j},{z_{imj}} \geqslant {\alpha _{im}} + {\rho _j} - 1, \hfill \\
  \forall i \in \mathcal{K},m \in \mathcal{M},j \in \mathcal{M}\backslash \left\{ m \right\}, {z_{imj}} \in \left\{ {0,1} \right\} \hfill \\ 
\end{gathered}
\end{equation}
Incorporating variables $z_{imj}$ to \eqref{eq:OPT-PF-MinSINRconFull}, and applying a simple re-arrangement of terms, we get:
\begin{multline}\label{eq:OPT-PF-MinSINRconFull-Ref1}
{g_{km}} \cdot \left( {L + 1} \right) - {g_{km}} \cdot \sum\limits_i {{\alpha _{im}}}  \geqslant \left( {\left( {{1 \mathord{\left/
 {\vphantom {1 p}} \right.
 \kern-\nulldelimiterspace} p}} \right) \cdot \sum\limits_i {{\alpha _{im}} \cdot {\alpha _{km}}} } \right) \cdot \theta  \\ + \left( {\sum\limits_i {\sum\limits_{j \ne m} {{g_{kj}} \cdot {z_{imj}}}  \cdot {\alpha _{km}}} } \right) \cdot \theta
\end{multline}

\paragraph{Step 2} With respect to the first term of the right-hand side of \eqref{eq:OPT-PF-MinSINRconFull-Ref1} we first linearize the user-association binary variable products, leveraging Technique~\ref{rem:BBP} as above. We introduce the auxiliary binary variables ${\left\{ {{v_{imk}}} \right\}_{i \in \mathcal{K},m \in \mathcal{M},k \in \mathcal{K}}}$ and the set of linear constraint expressions in \eqref{eq:AUX2} to replace each ${{\alpha _{im}} \cdot {\alpha _{km}}}$ product with a single ${{v_{imk}}}$ term.
\begin{equation}\label{eq:AUX2}
\begin{gathered}
  {v_{imk}} \leqslant {\alpha _{im}},{v_{imk}} \leqslant {\alpha _{km}},{v_{imk}} \geqslant {\alpha _{im}} + {\alpha _{km}} - 1, \hfill \\
  i \in \mathcal{K},m \in \mathcal{M},k \in \mathcal{K}, {v_{imk}} \in \left\{ {0,1} \right\} \hfill \\ 
\end{gathered} 
\end{equation}
  
\paragraph{Step 3} Likewise, the products ${z_{imj}} \cdot {\alpha _{km}}$ appearing in the second term of the right-hand side of \eqref{eq:OPT-PF-MinSINRconFull-Ref1}, are linearized with the introduction of binary auxiliary variables ${\left\{ {{u_{imjk}}} \right\}_{i \in \mathcal{K},m \in \mathcal{M},j \in \mathcal{M}\backslash \left\{ m \right\},k \in \mathcal{K}}}$ and the constraint expressions set \eqref{eq:AUX3}:
\begin{equation}\label{eq:AUX3}
\begin{gathered}
  {u_{imjk}} \leqslant {z_{imj}},{u_{imjk}} \leqslant {\alpha _{km}},{u_{imjk}} \geqslant {z_{imj}} + {\alpha _{km}} - 1, \hfill \\
  i \in \mathcal{K},m \in \mathcal{M},j \in \mathcal{M}\backslash \left\{ m \right\},k \in \mathcal{K}, {u_{imjk}} \in \left\{ {0,1} \right\} \hfill \\ 
\end{gathered}
\end{equation}

Based on linearization steps 2 and 3, \eqref{eq:OPT-PF-MinSINRconFull-Ref1} can be (exactly) rewritten as: 
\begin{multline}\label{eq:OPT-PF-MinSINRconFull-Ref2}
{g_{km}} \cdot \left( {L + 1} \right) - {g_{km}} \cdot \sum\limits_i {{\alpha _{im}}}  \geqslant \\ \left( {\left( {{1 \mathord{\left/
 {\vphantom {1 p}} \right.
 \kern-\nulldelimiterspace} p}} \right) \cdot \sum\limits_i {{v_{imk}} \cdot \theta } } \right) + \left( {\sum\limits_i {\sum\limits_{j \ne m} {{u_{imjk}} \cdot \theta } } } \right)
\end{multline}

For a known target SINR level $\theta$, \eqref{eq:OPT-PF-MinSINRconFull-Ref2} is linear with respect to the initial and new auxiliary variables, hence problem \eqref{eq:OPT-PF} has been transformed to a feasibility MILP. When a target SINR is not specified (as in our study), it is required to optimize over the SINR as well, and locate the maximum guaranteed SINR level for all UEs.  We observe that the non-linear terms in \eqref{eq:OPT-PF-MinSINRconFull-Ref2} involve products of a binary variable ($v_{imk}$ or $u_{imjk}$) and a bounded continuous variable (the SINR level $\theta$), which can be linearized according to Technique~\ref{rem:BCP}. The procedure is described in detail below.

%\footnote{Alternatively we could omit this linearization step and apply an iterative bisection search procedure, for which at each step a feasibility MILP problem is solved, for a given target SINR. Bisection search ultimately convergences to the maximum guaranteed level $\theta^*$ for a specified tolerance level, after typically 20-30 iterations (refer for example to \cite{GoAl13pimrc} and references therein for a complete description of the bisection search procedure and a relevant application.) 
%}
\paragraph{Step 4} Let $\mathcal{Q}$ a large positive number which is at least equal to the maximum expected achieved SINR level. We introduce two sets of (continuous) auxiliary variables $w_{imk}$ and $n_{ijmk}$ to replace the ${{v_{imk}} \cdot \theta }$ and ${{u_{imjk}} \cdot \theta }$ product terms respectively, along with the necessary linear constraint expression sets, \eqref{eq:AUX4a} and \eqref{eq:AUX4b}:
\begin{equation}\label{eq:AUX4a}
\begin{gathered}
  {w_{imk}} \leqslant \mathcal{Q}{v_{imk}},{w_{imk}} \leqslant \theta ,{w_{imk}} \leqslant \theta  - \left( {1 - {v_{imk}}} \right)\mathcal{Q}, \hfill \\
  {w_{imk}} \geqslant 0,i \in \mathcal{K},m \in \mathcal{M},k \in \mathcal{K} \hfill \\ 
\end{gathered}
\end{equation}
\begin{equation}\label{eq:AUX4b}
\begin{gathered}
  {n_{imjk}} \leqslant \mathcal{Q}{u_{imjk}},{n_{imk}} \leqslant \theta ,{n_{imjk}} \leqslant \theta  - \left( {1 - {u_{imjk}}} \right)\mathcal{Q}, \hfill \\
  {n_{imjk}} \geqslant 0,i \in \mathcal{K},m \in \mathcal{M},j \in \mathcal{M}\backslash \left\{ m \right\},k \in \mathcal{K} \hfill \\ 
\end{gathered}
\end{equation}
  
Combining all the reformulation steps, the original MINLP formulation converts to the following mixed integer linear form:
\begin{equation}\label{eq:OPT-MILP}
\normalsize
\left\{ \begin{gathered}
  \mathop {\text{maximize} } \theta {\text{ s}}{\text{.t}}{\text{. (2a),(2b),(6),(8),(9),(11),(12),}} \hfill \\
  {g_{km}}\left( {L + 1 - \sum\limits_i {{\alpha _{im}}} } \right) \geqslant \frac{1}{p}\sum\limits_i {{w_{imk}}}  + \sum\limits_i {\sum\limits_{j \ne m} {{n_{imjk}}} } , \hfill \\
  \forall i \in \mathcal{K},m \in \mathcal{M},j \in \mathcal{M}\backslash \left\{ m \right\},k \in \mathcal{K} \hfill \\ 
\end{gathered}  \right\}
\end{equation}

This problem can be solved leveraging powerful state-of-the-art ILP optimization tools, such as GUROBI, MOSEK or CPLEX, to acquire the optimal guaranteed SINR level and the user-association decision that realizes it. We remark that every step applied throughout this section provides an exact transformation of the original problem variables, and no approximation/relaxation penalty is induced. The price we have to pay is the increase in the number of variables and constraint expressions, caused by the introduction of auxiliary variables ($\left\{ {z,v,u,w,n} \right\}$) and necessary expressions linking the original and the auxiliary variables. However the elimination of non-linearity is more critical, since today's optimization tools are mature enough to handle large-scale MILPs in typical desktop computing platforms\footnote{As an example we report that a typical cluster instance involving 5 ANs and 10 UEs is solved in less than 5 secs using a laptop equipped with Intel Core i7-4710MQ CPU @ 2.50 GHz. using the GUROBI~\cite{GUROBI} solver called through CVX~\cite{CVX} in the MATLAB environment.}
%\cite{3GPP_R1133246},\cite{WiOs15}

\section{Results \& Discussion}\label{sec:Results}

\subsection{Simulation Setup}\label{sec:Results_SimSetup}

In this section we present the results of simulation experiments conducted with a two-fold perspective:
\begin{itemize}[leftmargin=10pt]
\item To highlight the benefits of joint and optimal load- and channel-based user-association in MM-UDNs, with respect to both achieved worse rate performance levels and the number of active ANs, for various infrastructure densification levels and Massive MIMO sizes;
\item To explore the optimal antenna element deployment in MM-UDNs, considering strategies that either concentrate elements on less ANs or distribute them in higher number of ANs, given a fixed overall antenna element budget. 
\end{itemize}

\begin{figure}[!t]
\centering
\includegraphics[scale=0.50]{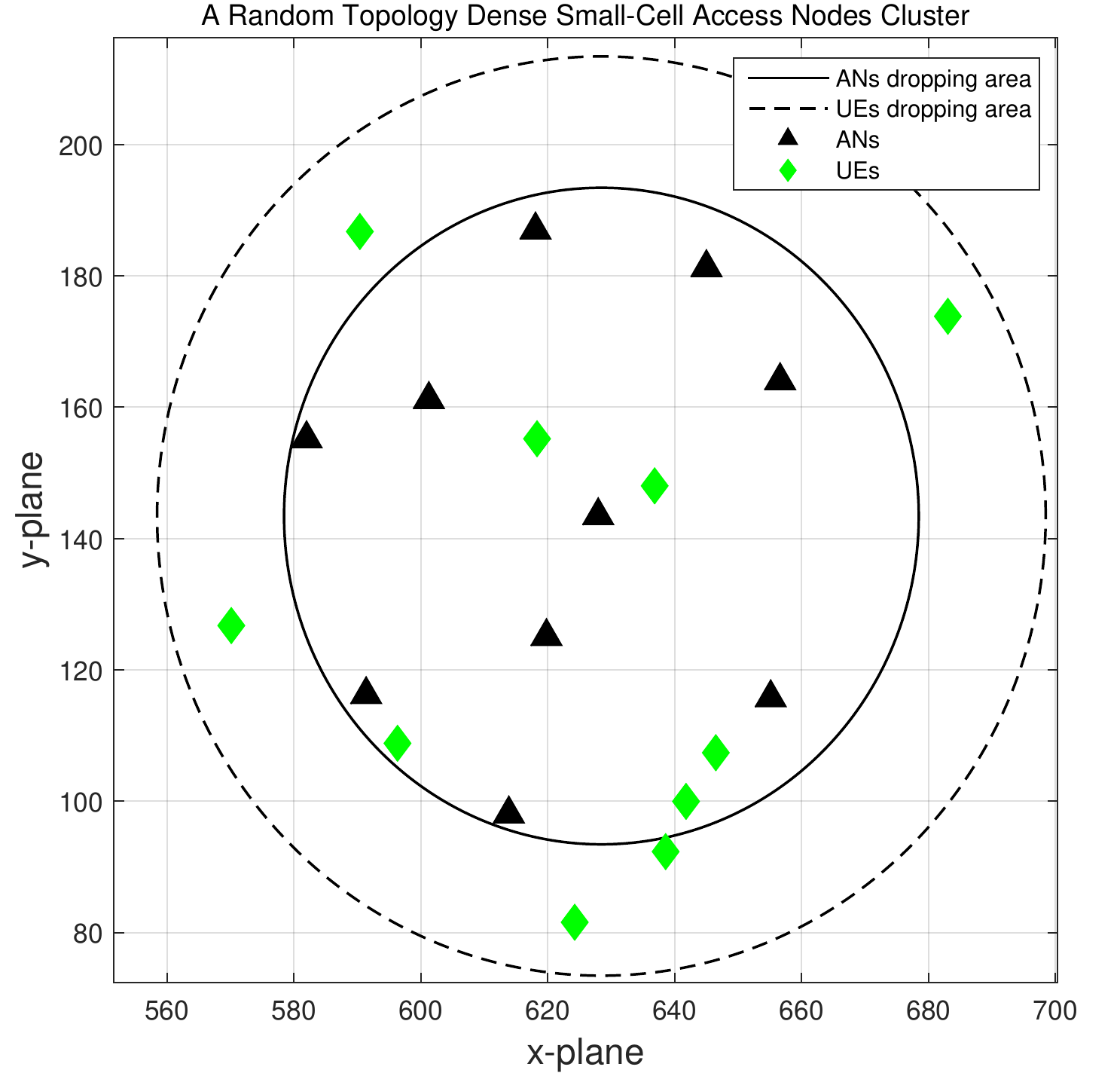} 
\caption{A small-cell cluster snapshot formed by 10 ANs and 10 UEs. Each cluster is dropped in a random hot-zone of an overlay macro-cell. We ignore inter-cluster interference. Within a single cluster ANs are dropped uniformly over a circular area with radius 50m and UEs over a co-centered area with radius 70m. ANs-UEs large-scale path-gains follow the ITU-UMi model as in \cite[A.1.2]{3GPP_36872}. Carrier frequency is set to 3.5 GHz and noise level to -174 dBm/Hz.}
\label{fig:SimulationSnapshot}
\end{figure}

We follow the simulation guidelines used within the 3GPP small-cell study item and particularly we consider the outdoor small-cell cluster scenario 2a~\cite[A.1.2]{3GPP_36872}. Figure~\ref{fig:SimulationSnapshot}
represents a snapshot generated based on settings reported in~\cite{3GPP_36872}. We extend the scenario by empowering small-cell ANs with massive number of antenna elements. We consider a fixed number of 10 UEs and vary the number of ANs from 2 to 10. Regarding power allocation, we assume a total cluster power budget which: i) is equally distributed to the deployed ANs (either active or inactive); ii) is independent from the number of ANs, hence with increased AN density the power level per AN ($p$) is reduced accordingly; iii) its absolute level is configured such that a specific spatially averaged interference-free SNR level is targeted. Two user-association strategies are considered, as discussed in Sec.\ref{sec:OPT_a} and Fig.\ref{fig:SystemSnapshot}:
\begin{itemize}[leftmargin=10pt]
\item The baseline channel-based approach, which associates each UE with the AN providing the maximum received signal power, and
\item The optimal channel- and load-based approach proposed in this paper.  For obtaining the optimal user-association decision for each system snapshot, we formulate the MILP problem in \eqref{eq:OPT-MILP} using the CVX modeling framework~\cite{CVX}, and use the state-of-the-art GUROBI package to solve it~\cite{GUROBI}. 
\end{itemize}
The results to be reported correspond to averaging over 200 independent system snapshots.

\subsection{Results}\label{sec:Results_SimRes}

\begin{figure}[!t]
\subfloat[{Worse UE Rate}]
{\includegraphics[width=0.47\textwidth]{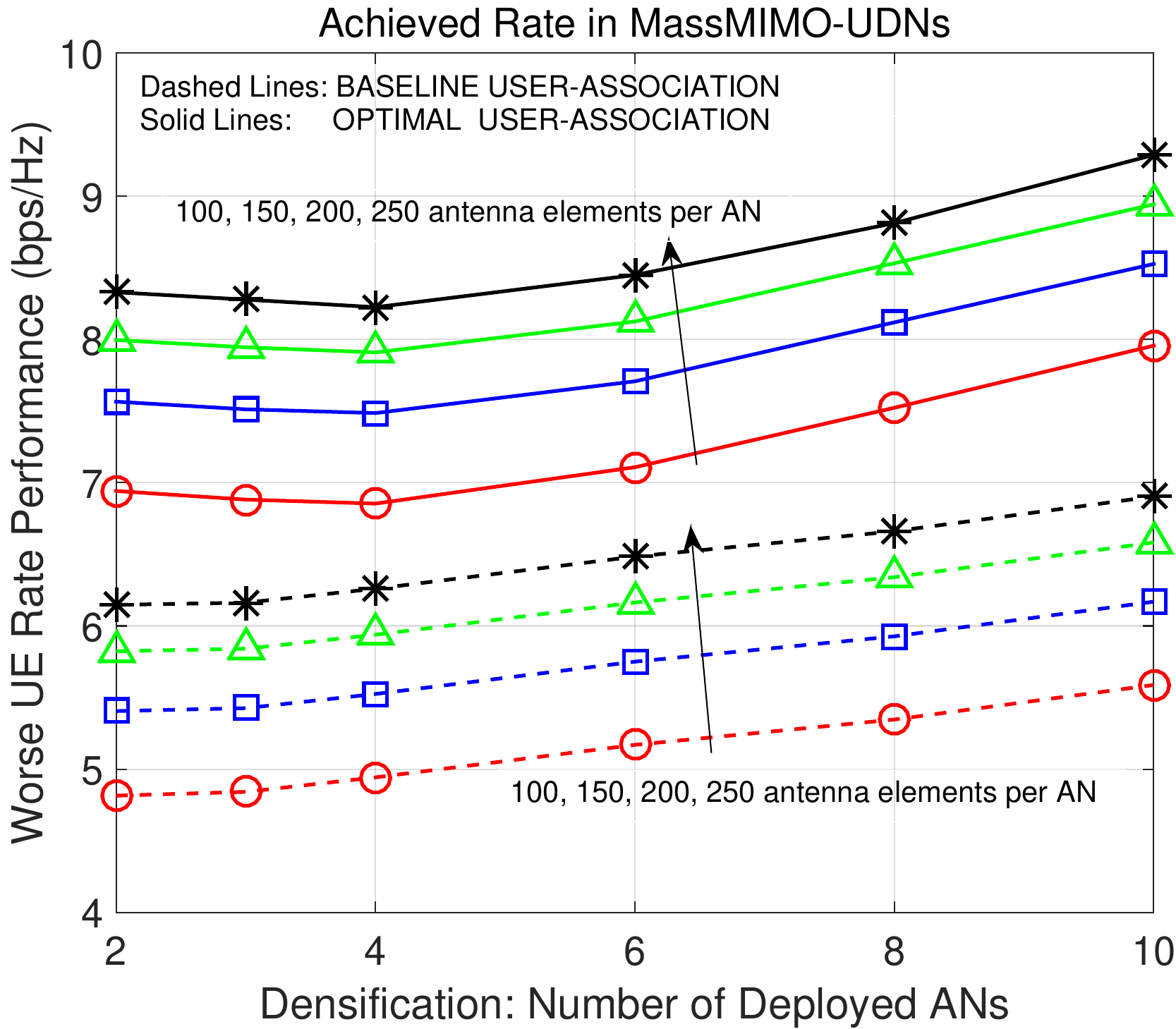}
\label{fig:Rate_vs_DensMM}} 
\hfill 

\subfloat[{Number of Active ANs}]
{\includegraphics[width=0.47\textwidth]{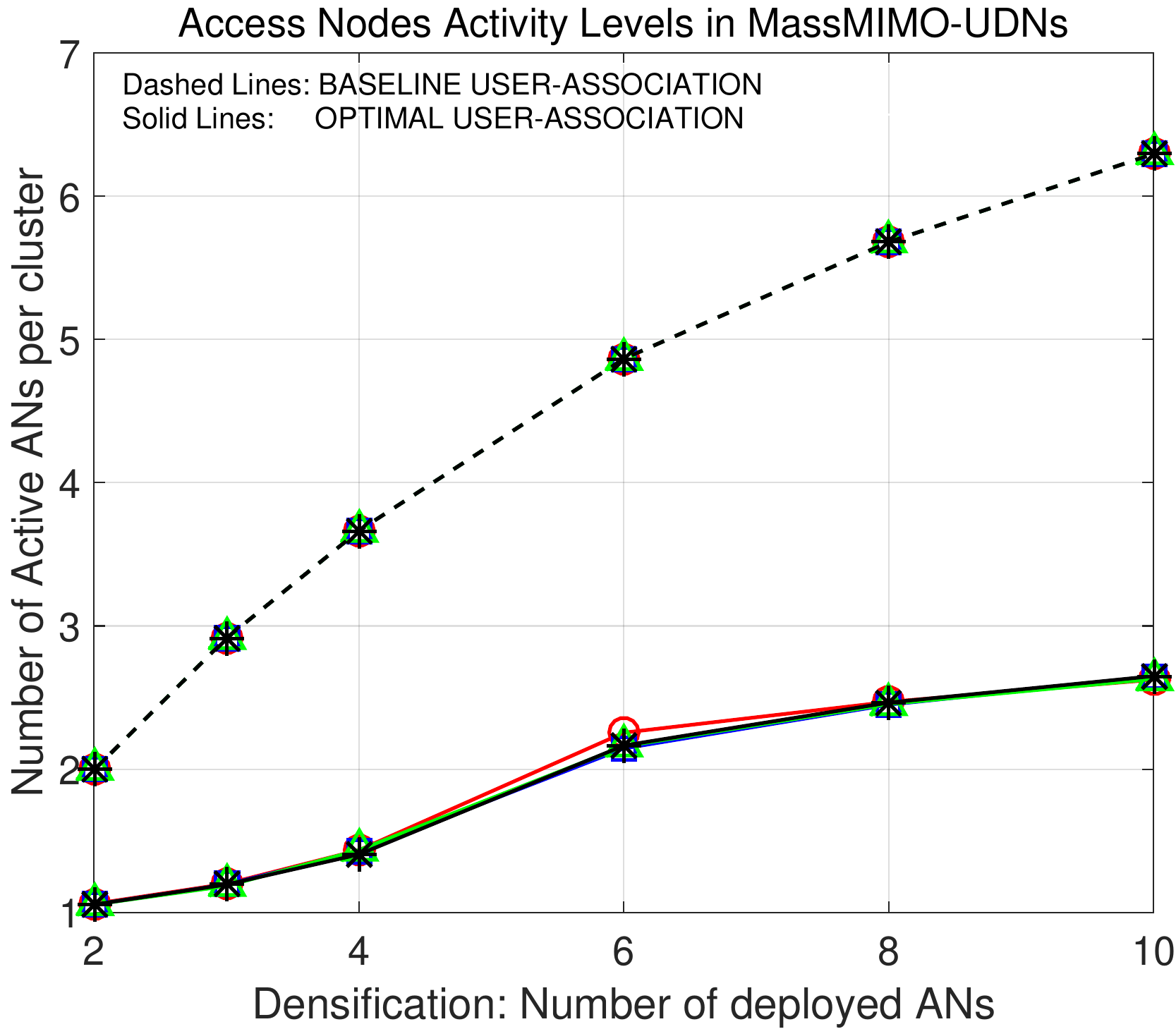}
\label{fig:NoOfActANs_vs_DensMM}}

\caption{The impact of network infrastructure densification (number of deployed ANs) and massive MIMO sizes (number of antenna elements per AN) for optimal and baseline user-association.}
\label{fig:Res_Set01}
\end{figure}

The first set of experiments demonstrates the importance of optimal user-association. An average target SNR of 30 dB is assumed. We compare the achieved worse UE rate performance for different infrastructure densification (number of deployed ANs within the considered small-cluster) and massive MIMO sizes (number of antenna elements per AN) for both user-association schemes. Figure~\ref{fig:Rate_vs_DensMM} shows that on average, 33-40\% rate performance gains are achieved by optimally selecting the serving AN per UE. The maximum gain is observed for the smallest MIMO size (100 elements). On top of the rate gains, Figure~\ref{fig:NoOfActANs_vs_DensMM} shows that the improved performance levels are achieved with a limited number of active ANs, on average with 50\% less active ANs than that of the baseline scheme. This means that load ``concentration" is a preferable strategy for the particular MM-UDN setups. We also observe that the benefit from increasing MM size is more prominent than that of infrastructure densification. As an example, for an MM size of 200 elements per AN, we observe a 12.5\% rate gain for a 2.5x infrastructure density increase (going from 4 to 10 ANs), whereas for the 6 ANs scenario, we observe a 20\% rate gain for a 2.5x MM size increase (going from 100 to 250 elements per AN).

\begin{figure}[!t]
\centering
\includegraphics[width=0.47\textwidth]{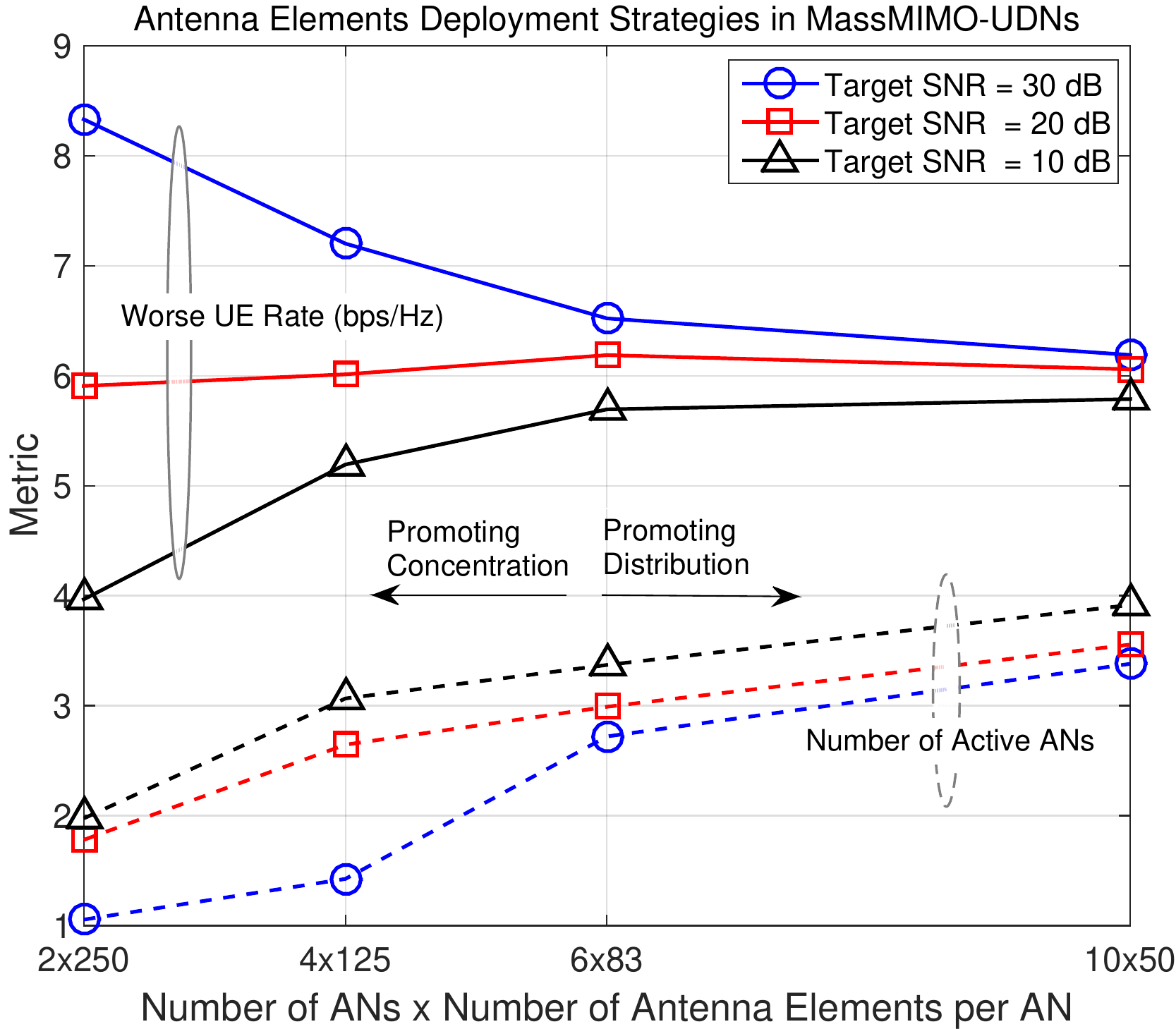} 
\caption{Impact of antenna elements deployment strategies to optimal worse UE rate and number of active ANs.}
\label{fig:ConcVsDistr}
\end{figure}

In the second set of experiments we consider a fixed amount of antenna elements (500) and study if it is preferable to concentrate these elements on sparse or dense AN deployments (resembling the macro vs small-cell network evolution dilemma). We apply the optimal user-association scheme of \eqref{eq:OPT-MILP} and examine the rate performance levels  as well as the number of active ANs supporting these levels for different element deployment strategies, namely having 2, 4, 6 or 10 ANs per cluster and 250, 125, 83 or 50 elements per AN respectively. Various transmission power levels are assumed corresponding to target SNR of 30, 20, and 10 dB. Results are presented in Figure~\ref{fig:ConcVsDistr}. It is deduced that the optimal strategy depends on the power budget. For high power availability (target SNR set to 30 dB) we observe a clear preference for concentrating antenna elements to a single or few ANs. This is justified by the fact that elements concentration allows to de-activate more ANs, reducing network interference levels, while the ANs are allocated enough power to support non-proximal links. In other words, proximity benefits offered by activating more ANs are not enough to counterbalance the excess interference ``leakage". For limited power budgets (target SNR set to 10 dB) the distribution policy is optimal, since proximity benefits are critical for dealing with limited received signal levels. Regarding the number of active ANs supporting the optimal rates, we observe that it decreases as the target SNR is higher, since the optimal network operation point moves from the noise-limited to the interference-limited regime.

\section{Conclusion}\label{sec:Concl}
Increasing the number of access elements per user is regarded as a major evolution path towards the next (5th) generation of wireless networks. This could be achieved by heavy densification of access nodes (Ultra-Dense Networks) and/or heavy increase in MIMO dimensions (Massive MIMO). Within this framework, we examined the problem of optimal channel- and load-based user-association. We reformulated the original computationally-intractable problem to a more plausible form, allowing us to acquire the optimal association decisions with reasonable complexity. We applied the optimization model to typical dense small-cell setups and demonstrated that optimal user-association significantly enhances rate performance levels and offers additional benefits by reducing the number of active access nodes. In addition we explored alternative strategies for optimally deploying the massively available number of access elements to ultra-dense networks.

\section{Acknowledgement}
This work has been performed in the context of the ARTCOMP PE7(396) ``\emph{Advanced Radio Access Techniques for Next Generation Cellular NetwOrks: The Multi-Site Coordination Paradigm}'' research project, implemented within the framework of Operational Program “Education and Lifelong earning”, co-financed by the ESF and the Greek State.

\bibliographystyle{IEEEtran}
\bibliography{IEEEabrv,ARTCOMP-bib}

\end{document}